\documentclass[aps,prl,reprint,groupedaddress]{revtex4-1}
\usepackage{graphicx}

\newcommand{\um}{\mu \mathrm{m}}
\newcommand{\nm}{\mathrm{nm}}

\newcommand{\vet}{\mathbf}
\newcommand{\alzcq}{\mathrm{Al}_{0.5}\mathrm{Ga}_{0.5}\mathrm{As}}
\newcommand{\micron}{\mu \mathrm{m}}
\usepackage{amsmath}

\begin{document}
\title{Photonic bands, superchirality, and inverse design of a chiral minimal metasurface}

\author{Simone Zanotto}
\affiliation{Istituto Nanoscienze - CNR and Laboratorio NEST, Scuola Normale Superiore, Piazza San Silvestro 12, 56127 Pisa, Italy}
\email{simone.zanotto@nano.cnr.it}
\author{Giacomo Mazzamuto}
\affiliation{European Laboratory for Non Linear Spectroscopy, via N. Carrara 1, Sesto Fiorentino, Firenze 50019, Italy}
\affiliation{CNR-INO Sesto Fiorentino, via Nello Carrara 1, Firenze 50019, Italy}
\author{Francesco Riboli}
\affiliation{European Laboratory for Non Linear Spectroscopy, via N. Carrara 1, Sesto Fiorentino, Firenze 50019, Italy}
\affiliation{CNR-INO Sesto Fiorentino, via Nello Carrara 1, Firenze 50019, Italy}
\author{Giorgio Biasiol}
\affiliation{Istituto Officina dei Materiali CNR, Laboratorio TASC, Basovizza (TS) - Italy}
\author{Giuseppe C. La Rocca}
\affiliation{Scuola Normale Superiore and CNISM, Piazza dei Cavalieri 7, 56126 Pisa, Italy}
\author{Alessandro Tredicucci}
\affiliation{Istituto Nanoscienze - CNR and Laboratorio NEST, Scuola Normale Superiore, Piazza San Silvestro 12, 56127 Pisa, Italy}
\affiliation{Dipartimento di Fisica ``E. Fermi'', Università di Pisa, Largo Pontecorvo 3, 56127 Pisa, Italy}
\author{Alessandro Pitanti}
\affiliation{Istituto Nanoscienze - CNR and Laboratorio NEST, Scuola Normale Superiore, Piazza San Silvestro 12, 56127 Pisa, Italy}
\date{\today}
\begin{abstract}
Photonic band structures are a typical fingerprint of periodic optical structures, and are usually observed in spectroscopic quantities such as transmission, reflection and absorption. Here we show that also the chiro-optical response of a metasurface constituted by a lattice of non-centrosymmetric, L-shaped holes in a dielectric slab shows a band structure, where intrinsic and extrinsic chirality effects are clearly recognized and connected to localized and delocalized resonances. Superchiral near-fields can be excited in correspondence to these resonances, and anomalous behaviors as a function of the incidence polarization occur. Moreover, we introduce a singular value decomposition (SVD) approach to show that the above mentioned effects are connected to specific fingerprints of the SVD spectra. Finally, we demonstrate by means of an inverse design technique that the metasurface based on an L-shaped hole array is a \textit{minimal} one. Indeed, its unit cell geometry depends on the smallest number of parameters needed to implement arbitrary transmission matrices compliant with the general symmetries for 2d-chiral structures. These observations enable more powerful wave operations in a lossless photonic environment. 
\end{abstract}

\maketitle

\section{Introduction}
Electromagnetic fields which carry chirality -- in their simplest form, left- and right-circularly polarized plane waves -- deserve huge interest as they interact with matter chirality, enabling for instance to discriminate enantiomers in chemistry, which are ultimately connected with key features of living organisms. Indeed, many biomolecules have a specific handedness, and it is not yet clear why nature has decided to go in that precise direction \cite{LoughWainer}. From a more application-oriented point of view, the pharmaceutical industry constantly seeks for effective methods to discriminate stereoisomers, an application where chiral light-matter interaction could prove useful. 

To date, the most common technique to prepare and analyze chiral light is to employ birefringent plates and linear polarizers that convert light to and from linear polarization, as the technology of direct sources and detectors of chiral light is still in its infancy \cite{LobanovPRB2015, DyakovPRB2018, KonishiPRL2011, SollnerNatNano2015}. 
Last advances in nanotechnology are however revolutionizing chiral optical devices \cite{PfeifferPRAppl2014, ZhaoNatComm2012, ZhaoNatComm2017, HentschelSciAdv2017, KongAM2018, AsadchyNanophotonics2018, TulliusJACS2015, SchaeferlingACS2016, VazquezGuardadoPRL2018, ShaltoutOptica2015, yin2015active, Eismann2018, Fang2019, Zektzer2019, kong2018photothermal, mohammadiACS2018}. Far- and near-field chiral electromagnetic responses have been indeed observed in a variety of artificially structured systems, where the shape of the machined elements must have a three-dimensional character if geometric and electromagnetic chirality has to be attained in its most rigorous form, because of the requirement of the absence of any mirror symmetry plane  \cite{PlumZheludev, FernandezCorbatonPRX2016, GarciaSantiagoOL2017}. Several proposed structures hence rely on volumetric fabrication techniques, which however suffer from either scarce throughput or limited flexibility. Thus, less demanding fabrication technologies -- i.e., planar technologies -- may also be employed, as witnessed by some reports \cite{KuwataPRL2005, WuNatComm2014, YePRAppl2017}. For instance, a dielectric film with a non-centrosymmetric partially etched planar pattern was proposed as a simple gateway towards strong chiro-optical phenomena \cite{ZhuLSA2018}. 
Nonetheless, there is a wide interest in developing subwavelength-patterned high-index dielectrics, both for applications and for fundamental research: from one side, they enable the synthesis of flat lenses, polarimeters, spectrometers, nonlinear components and computer-generated holograms \cite{ZhaoLSA2018, ArbabiNatNano2015, KarakasogluOE2018, ZhuAPL2017, BalthasarPRL2017, Li2017}; from another side, they exhibit a variety of intriguing phenomena such as Fano lineshapes, perfect forward scattering, geometric phase effects, and bound states in the continuum resonances \cite{Limonov2017,  Fu2013, Person2013, Kim2015, Hsu2016}. 

Most of these effects arise from complex electromagnetic behaviors whose essence can be grasped understanding the interplay between localized and delocalized resonances. In other words, these systems often rely on the co-presence, and on the competition, between guided wave phenomena and antenna-like responses, as highlighted by the prototypical transition between the guided mode filter regime, the photonic crystal regime, and the independent-particle Mie-scattering regime \cite{CollinRPP2014}. In this work we report on the observation of photonic bands in a chiral metasurface, highlighting that the chiral response shows fingerprints of both guided wave and locally resonant phenomena. The object under investigation is the simplest conceivable 2d-chiral dielectric metasurface: a slab perforated with L-shaped holes \cite{MenzelPRA2010}. Although it is not a 3d-chiral geometric object, its finite thickness allows it to implement interesting chiral electromagnetic features such as chiro-optical far-field response at normal incidence and the excitation of superchiral near-fields with incident unpolarized light. The metasurface response at normal incidence will be analyzed by means of a singular-value decomposition approach that reveals the operational capabilities of any 2d-chiral photonic device. Moreover, we will show that the extremely simple L-shaped structure is also a \textit{minimal} one: by tuning a small number of geometrical parameters, it is possible to access a very wide set of the transmission matrices characteristic of a general 2d-chiral metasurface. In this sense, we approached the solution of the inverse design problem of 2d-chiral metasurfaces \cite{Molesky2018}.

\section{Chiral response at normal incidence}
The metasurface under investigation is illustrated in Fig.~1a. 
\begin{figure}[htbp]
\includegraphics{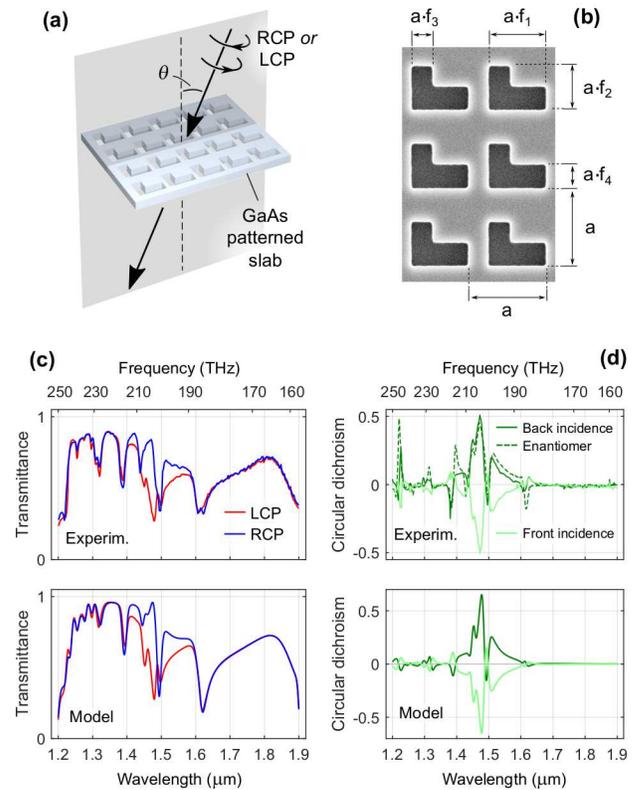}%
\caption{\label{fig1} (a) Schematic of the metasurface and of the optical measurement. RCP and LCP indicate, respectively, right- and left-circularly polarized light. (b) Scanning electron micrograph of the metasurface with its geometrical parameters. (c) Polarization-resolved transmittance spectra, both from experiment and from numerical modeling. (d) Circular dichroism spectra (i.e., difference between LCP and RCP transmittances) measured on the metasurface represented in (b) from both front side and back side, and from the front side of a metasurface having the enantiomeric pattern. Data in panels (c) and (d) are obtained at $\theta = 0$.}
\end{figure}
It consists of a 220 nm thick gallium arsenide membrane, patterned with L-shaped holes arranged over a square lattice. The geometric parameters of the holes are reported in Fig.~1b, superimposed to the scanning electron microscope (SEM) image of a fabricated sample. While in a forthcoming analysis the parameters $a$, $f_1 \ldots f_4$ will be allowed to vary, the first part of the article deals with a specific choice of parameters, which correspond indeed to the SEM image. In detail, we have $a = 1134\ \mathrm{nm}$, $f_1 = 0.76$, $f_2 = 0.58$, $f_3 = f_4 = 0.327$. The sample fabrication process, whose details are reported in the Supplementary Materials, allows to obtain a frame-supported membrane, freely accessible from both sides to perform optical measurements with a moderately focused ($\approx 50\ \um$ spot) near-infrared (1 -- 2 $\um$ wavelength) beam. The beam originates from a supercontinuum source, filtered by means of an acousto-optic tunable filter yielding a spectral bandwidth of $\approx 2\ \nm$. The system is controlled through an automated software developed in C++. The polarization state is prepared with a Glan-Taylor polarizer followed by a $\lambda/4$ superachromatic waveplate; no polarization analysis is performed on the beam transmitted after the sample. The sample is mounted on a rotating support that enables to measure the angularly-resolved transmittance, as illustrated in Fig.~1a. A first set of measurements has been however collected at normal incidence ($\theta = 0$). The experimental data are reported in Fig.~1c, and are compared with the outcome of a numeric model (rigorous coupled wave analysis, RCWA -- see the Supplementary Materials for details). The spectra consist of a series of quite narrow dips, some of them having different shapes and depths depending on the polarization state of the incident light. From the transmittances for left (right) circularly polarized light, respectively $\mathcal{T}_L$ ($\mathcal{T}_R$), the transmission circular dichroism can be defined as $\mathit{CD} = \mathcal{T}_L - \mathcal{T}_R$: this quantity is plotted in Fig.~1d. Here, the traces labeled \textit{front incidence} and \textit{back incidence} have been collected from the same sample, but illuminated from either the top surface or the bottom surface. The trace \textit{enantiomer} derives instead from a second sample, which differs from the first one solely for the L-shaped hole sense of chirality at fabrication stage: the second sample, when observed from the front side, is described by the same parameters of the first sample exception made for the exchange $f_1 \leftrightarrow f_2$. The good matching between the CD spectrum of the enantiomer with that of the original sample's back side, as well as the fact that the front and back side CD are opposite to each other, indicates that the objects under investigation are almost perfectly 2-d chiral. Indeed, while in principle a through-hole perforated homogeneous membrane has a mirror plane parallel to its surface, fabrication imperfections may have lead the actual sample to deviate from this ideal characteristic. The data instead indicate that the two samples are essentially the same object, from both geometric and electromagnetic points of view. 

The data reported so far indicate that a 2d-chiral metasurface with a very simple design exhibits strong circular dichroism in a narrow band close to the optical telecommunication window. It should be emphasized that the observed CD is not accompanied by absorption; rather, it originates from a redistribution of the incident energy among the transmitted and reflected beams (no diffraction is present thanks to the subwavelength dimensions of the pattern). In this sense, the presented object is rather a beam splitter sensitive to the circular polarization. However, from the point of view of an observer who can only access the transmitted beam, the effect is indistinguishable from dichroic absorption induced CD. Hence, this effect can be employed to mimick the CD of target objects such as, for instance, naturally occurring molecules.

\section{Chiral response at oblique incidence and band structure}
In the transmission spectra reported in Fig.~1 a rich structure can be noticed, whose physical origin deserves attention \textit{per se} and in view of applications. In order to get further insight into the nature of the resonances leading to CD peaks, a powerful method is to measure the metasurface transmittance at different angles of incidence (i.e., for different orientations of the incident wavevector). Angularly-resolved measurements have been indeed recently employed to reveal special features of polarization phenomena like magneto-optic effects in quasi-ordered structures \cite{Kalish2018} and asymmetric transmission in low-symmetry plasmonic hole arrays \cite{Arteaga2014}. This approach allows to distinguish between angularly dispersive and non-dispersive resonances, which originate from different physical phenomena: guided mode resonances and antenna-like resonances, respectively. In Fig.~2 a-b we report the CD mapped in the frequency-wavevector space, where the fingerprints of the above cited mechanisms better stand out.
\begin{figure*}[h!]
\centering
\includegraphics{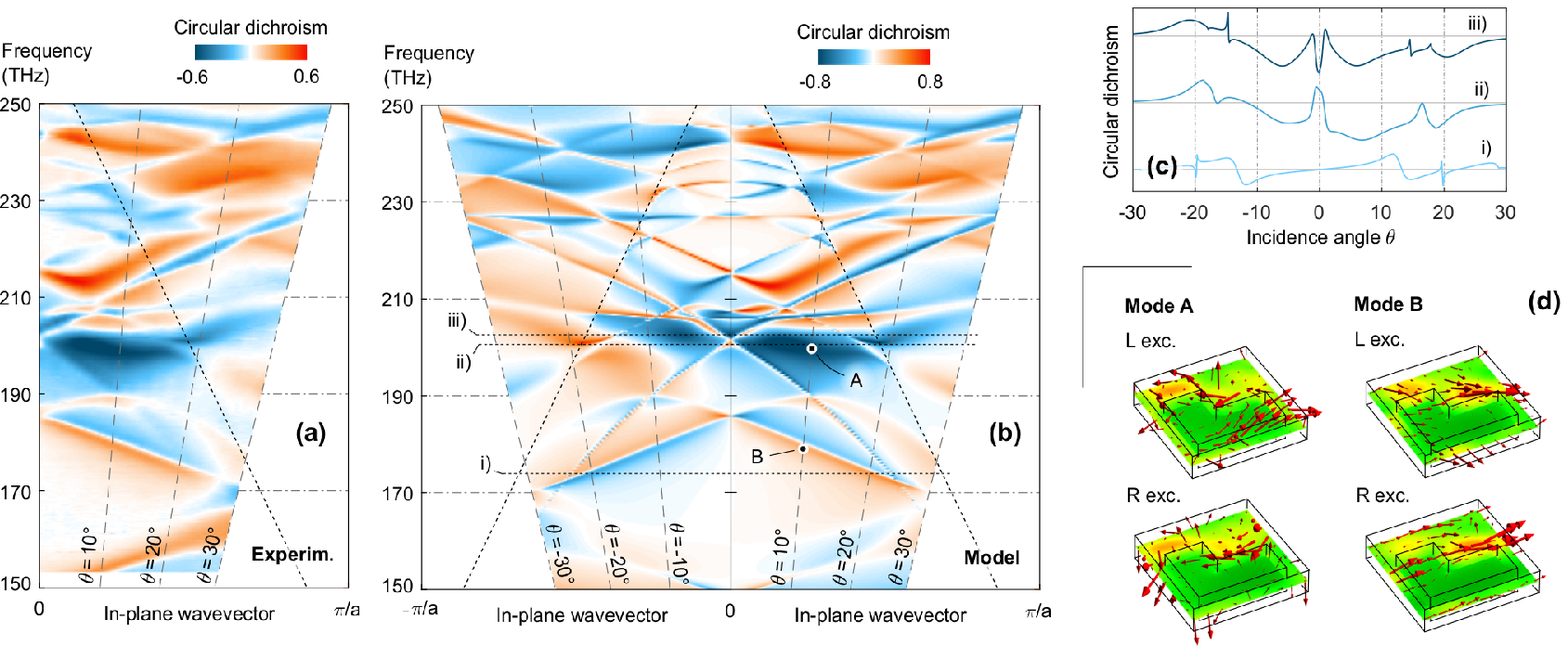}%
\caption{\label{fig2}  Measured (a) and modeled (b) transmission circular dichroism (CD) band structure of the metasurface. A rich pattern originating from the interplay between guided mode resonances and antenna-like resonances is observed. The diagonal short-dash lines represent the diffraction threshold: below those lines the sample operates in the proper metasurface regime. By tuning the incidence angle $\theta$ the first Brillouin zone is probed along a high-symmetry direction (see also Fig.~1 (a)). Notice that the CD does not fulfil the relation $CD\theta = - CD(-\theta)$, consistently with the expectation for an intrinsically 2d-chiral metasurface. Panel (c) highlights this behavior, for three optical frequencies identified in panel (b). Panel (d) illustrates the electromagnetic energy density (color map: green = low, orange = high) and the Poynting vector (red arrows) inside the metasurface unit cell for modes A and B (see panel (b) for their definition) and for both circular polarization states of the illuminating beam. Notice that in mode B the Poynting vector follows a well-defined direction, witnessing the presence of a traveling wave, while in mode A it ``winds up'' around the L-shaped inclusion, indicating a localized resonance.}
\end{figure*}
Here, the in-plane wavevector is the projection of the incident light wavevector on the metasurface plane. In the experiment only positive angles have been studied, while the model data are available for both positive and negative angles. The good matching between experimental and model data allows to acquire confidence in the model as a whole, which shows interesting features. First, strongly dispersive bands, with positive and negative slope, suggest that the metasurface optical response has important contributes from guided mode resonances (also known as quasi-guided modes); this picture is supported by an empty-lattice band-folding model illustrated in the Supplementary Material. These resonances are responsible for the narrow dips observed in the transmission spectra and for the sharp oscillations in the CD spectra. 

Second, by comparing the CD observed at opposite angles, it can be noticed that in the region at the center of the map ($-20^{\circ} < \theta < 20^{\circ}$, $190\ \mathrm{THz} < \nu < 210\ \mathrm{THz}$) the CD is not symmetric with respect to the exchange $\theta \leftrightarrow -\theta$. The effect can be better noticed looking at the curves in Fig.~2c, where the CD as a function of the incidence angle has been reported for three fixed values of light frequency (labels i--iii in the figure). We attribute this effect to the presence of a spectrally and angularly broad chiral resonance, which we will refer to as L-hole resonance. This resonance has a behavior that contrasts in two ways that of the guided mode resonances: first, the broad angular response of the L-hole resonance suggests that it has a spatially localized nature, as opposed to the delocalized, traveling-wave nature of the guided mode resonances (which instead follow a dispersive energy-wavevector curve). Second, it entails an intrinsic chiral response: while in the outer regions of the map, where the L-hole resonance is inactive, the CD displays an \textit{extrinsically chiral} character ($CD(\theta) = - CD(-\theta)$), in the central region of the map an \textit{intrinsically chiral} character emerges ($CD(\theta) \neq - CD(-\theta)$) \cite{Plum2009}. 

Seeking for further confirmation of this view, we calculated (see the Supplementary Materials for details) the optical near fields of the metasurface unit cell, upon the illumination conditions that excite the two types of resonances. The data are plotted in Fig.~2d, where both energy density (colormap) and Poynting vector (arrows) are reported. While the energy density distributions do not show remarkable behaviors, the Poynting vector field shows interesting properties that are consistent with the picture sketched above. When the metasurface is illuminated with the energy-wavevector pair labeled as A in Fig.~2b, the Poynting vector in the unit cell has a strongly inhomogeneous distribution: for left polarized excitation, it even ``winds up'' around the L-shaped hole. For right-polarized excitation the Poynting vector has a different, yet still irregular, distribution. We interpret these irregular distributions as arising from the L-hole resonance: a localized resonance dominated by multipoles, similarly to the reports of Ref.~\cite{ZhuLSA2018}. On the contrary, when the metasurface is illuminated with the energy-wavevector pair labeled B, the Poynting vector shows much more regular distributions, typical of a guided mode that propagates parallel to the slab. Noticeably, the direction of the Poynting vector is opposite with respect to that of the in-plane wavevector, consistently with the negative dispersion of the photonic band where point B lies on.

\section{Superchiral near-fields}
Besides far-field angularly-dispersive chiro-optic response, which is of interest -- for instance -- for filters, holograms, and multiplexers, near-field optical chirality plays a fundamental role in the interaction with chiral matter. First relegated to the role of pure mathematical curiosity, electromagnetic (e.m.) chirality is now regarded as a crucial quantity to be taken care of when the optical detection of chiral molecules is under investigation \cite{VazquezLozano2018, Alpeggiani2018, Tang2010, Tang2011}. The ability of a photonic structure to enhance the e.m.~chirality is quantified by the \textit{normalized e.m.~chirality}, defined as $\hat{C} = -c~ \mathrm{Im} (\vet{E}^* \cdot \vet{B})/|\vet{E}_{\mathrm{inc}}|^2$, where $c$ is the speed of light, $\vet{E}$ and $\vet{B}$ are the fields at the point of interest, and $\vet{E}_{\mathrm{inc}}$ is the electric field vector of the incident plane wave. In the absence of any photonic structure, the plane wave would freely propagate and one would have $\hat{C} = 0$ for linear polarization and $\hat{C} = \pm 1$ for right (left) circular polarization. When the incident wave impinges instead on a photonic structure, the near-fields excited in its vicinity may display a significantly different value of  $\hat{C}$: for instance, one may have $\hat{C} \neq 0$ also if the incident light is linearly polarized, or it is possible that $|\hat{C}| \gg 1$ (\textit{superchiral field}). In Fig.~3 we show that the 2d-chiral L-shape patterned membrane has a strong near-field e.m.~chirality, also at normal incidence, and also when considering the spatial average of $\hat{C}$ over a plane placed in closed proximity of the metasurface (figure inset).
\begin{figure}[htbp]
\begin{center}
\includegraphics[width = 6.5 cm]{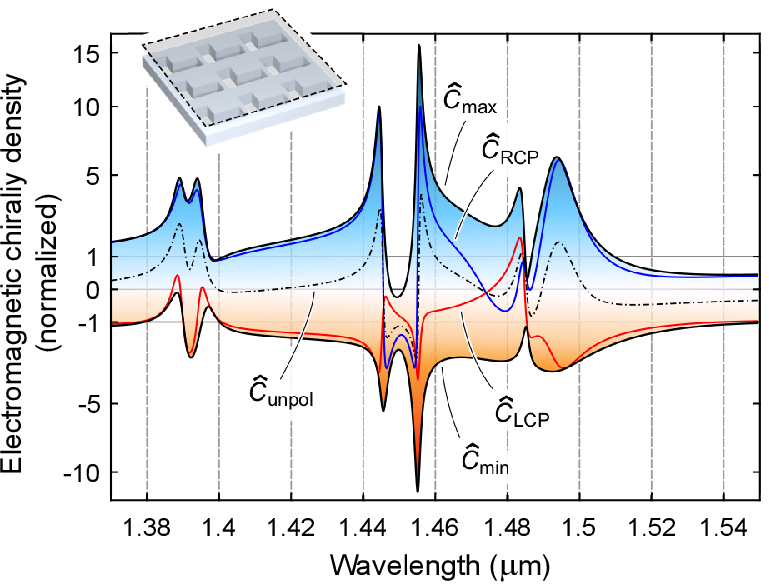}%
\end{center}
\caption{\label{fig3} Simulated electromagnetic chirality density $\hat{C}$ calculated by averaging the local chirality density on a plane above the metasurface (see the figure inset and the main text for details). The different traces report the values of $\hat{C}$ that can be attained upon different incident polarization conditions: left- and right-circular (LCP and RCP) and unpolarized (\textit{unpol}). $\hat{C}_{\mathit{max}}$ and $\hat{C}_{\mathit{min}}$ are the absolute maximum and minimum attainable values of $\hat{C}$; they correspond, in general, to an elliptical polarization state of the incident light. Normal incidence is assumed. Free-space propagating RCP and LCP have $\hat{C} = \pm 1$, respectively. }
\end{figure}
In the graph, the colored area indicates the accessible values of $\hat{C}$ for the present structure. Superchiral behavior occurs at wavelengths corresponding to the spectral features observed in the far-field transmission (see Fig.~1c). Noticeably, maximum (minimum) values of $\hat{C}$ are not always achieved upon right (left) circularly polarized illumination, as it can be observed from the corresponding traces (blue and red curves, respectively). Even more curiously, at the wavelength of $1.48\ \um$ a reversal occurs ($\hat{C}_{\mathrm{RCP}} < \hat{C}_{\mathrm{LCP}}$). Furthermore, a strongly anomalous value $\hat{C}_{\mathrm{max}} < 0$ is observed at $\lambda = 1.45\ \um$. We attribute these anomalies to the influence of the chiral localized resonance, ``mode A'', discussed above, whose interplay with the narrower guided mode resonances leads to such features. In the considered spectral region other interesting phenomena take place. First, the metasurface displays $\hat{C}\neq 0$ also when it is illuminated with unpolarized light, as illustrated by the trace $\hat{C}_{\mathrm{unpol}}$. Second, the definite chirality of the metasurface enables an asymmetry between the extremal values of $\hat{C}$, which do not fulfil the equality $\hat{C}_{\mathrm{max}} = -\hat{C}_{\mathrm{min}}$. In essence, an appropriate choice of wavelength allows to engineer the near-field chiral response to a large extent, as the device response breaks certain conventional rules concerning e.m.~chirality.

\section{Singular value decomposition analysis of the metasurface operation}
\label{sect:svd}
We now move back to the far-field metasurface response, in the attempt to understand in a comprehensive way its amplitude, phase and polarization response. This analysis relies on the properties of the \textit{T}-matrix (transmission matrix, or Jones matrix), where the device response concerning light transmission is completely encoded\footnote{We recall that we are limiting ourselves to the regime where the wavelength is larger than the pattern periodicity, i.e.~to the \textit{metasurface regime}, where no diffraction takes place.}. The $T$-matrix can be written over different bases; the most common choices being that of linearly or circularly polarized waves. In the following we will identify the $T$-matrix written in these bases as $T_{\mathrm{L}}$ and $T_{\mathrm{C}}$, respectively. For instance, $T_{\mathrm{C}}$ operates over Jones vectors whose elements are the right- and left-handed circularly polarized components of incident/transmitted light: $\vet{t} = T_{\mathrm{C}} \vet{i}$, with $\vet{t} = (t_R, t_L)$ and $\vet{i} = (i_R, i_L)$.

The form of $T$, and hence the possible operations that a metasurface can implement, are dictated by geometrical symmetry properties of the pattern lattice and unit cell. For instance, the L-shaped hole structure belongs to the more general category of objects that have a single mirror symmetry plane, usually identified by $M_{x,y}$ where the $x-y$ plane is that of the metasurface. Under normal incidence, the $T_L$ matrix of such metasurface is symmetric: $T_{\mathrm{L}} = T_{\mathrm{L}}^{t}$ \cite{MenzelPRA2010}. However, to get a more insightful vision into the metasurface operation we found it useful to focus on the properties of $T_{\mathrm{C}}$, in particular to its singular value decomposition (SVD). The SVD is an algebraic operation that reveals the structure of non-unitary linear operators, similarly to what eigenvalue decomposition (ED) does about unitary operators; a short recall of the properties of the SVD are given in the Supplementary Material. In the case under analysis the SVD is better suited with respect to ED, since the $T$-matrix is in general not unitary. Indeed, a dielectric metasurface is lossless in its full response (reflection plus transmission), but it appears lossy as the sole transmission is considered. The SVD of $T$ naturally describes the energy-handling behavior of the metasurface, since the square of its two singular values are the maximum and minimum wave intensity that can be transmitted by the metasurface (assuming unit intensity incident beam)  \cite{Ge2017}. Thus, our approach goes beyond what was reported in past studies on planar chiral metamaterials which did not rely on the SVD \cite{BaiPRA2007, BaiOE2009}, and gives a complementary viewpoint with respect to other decomposition methods (pseudopolar and integral decomposition, \cite{Arteaga2009, Ossikovski2019}) employed in polarization optics.

Algebraic manipulations (see the Supplementary Material) leveraging on the symmetry of $T_{\mathrm{L}}$ imply that the unitary matrices $V$ and $W$ appearing in the SVD of $T_{\mathrm{C}}$, i.e.,  $T_{\mathrm{C}} = V \Sigma W^{\dagger}$, must fulfil $\bar{V}W = \left( \begin{array}{cc} 0 & 1 \\ 1 & 0 \end{array} \right)$. We recall that $\Sigma$ is a diagonal matrix containing the singular values $\sigma_{1,2}$, and that the notation $\bar{V}$ indicates the element-wise matrix complex conjugate. The constraint on $V$ and $W$ implies that these matrices can be parametrized in the following insightful form:
\[
V = -i e^{i\tilde{\phi}/2} D_1 O D_2, \qquad W = e^{-i\tilde{\phi}/2} D_1' O' D_2'
\]
where $D_1 = D_1' = \left( \begin{array}{cc} e^{i \psi} & 0 \\ 0 & e^{-i \psi}  \end{array} \right)$, $O = \left( \begin{array}{cc} \sin \theta & \cos \theta \\ -\cos \theta & \sin \theta  \end{array} \right)$, $O' = \left( \begin{array}{cc} \cos \theta & \sin \theta \\ -\sin \theta & \cos \theta  \end{array} \right)$, $D_2 = \left( \begin{array}{cc} -i e^{-i \Delta} & 0 \\ 0 & i e^{i \Delta}  \end{array} \right)$, $D_2' = \left( \begin{array}{cc} e^{i \Delta} & 0 \\ 0 & e^{-i \Delta}  \end{array} \right)$. With this notation one has $T_{\mathrm{C}} = D_1 F D_1^{\dagger}$, which means that the metasurface performs a ``core'' operation, described by $F$, transformed by the basis change identified by the unitary matrix $D_1$. Noticeably, $D_1$ is the matrix that describes, in the circular polarization basis, a rotation of the reference system about the axis perpendicular to the metasurface plane. In other words, the operation of an $M_{x,y}$-symmetric metasurface can be summarized in the operation encoded by the operator $F = -ie^{i\tilde{\phi}} O D_2 \Sigma D_2'^{\dagger} O'^{t}$, that however acts in a rotated coordinate system. From this fact it follows that, in order to specify the operation of a $M_{x,y}$-symmetric metasurface, it is sufficient in essence to specify its $F$: indeed, the full space of transmission matrices can be saturated by simply rotating the metasurface about the axis perpendicular to the metasurface plane.

We now devote our attention to understanding the meaning of the parameters entering the matrices above. To this end one should explicitly describe the operation of $T_\mathrm{C}$ over a generic incident field vector $\vet{i}$. By decomposing it over the basis defined by the columns $\vet{w}^{(1,2)}$ of the unitary matrix $W$, one has $T_\mathrm{C} \vet{i} = T_\mathrm{C} \left( i_1 \vet{w}^{(1)} + i_2 \vet{w}^{(2)} \right) = \sigma_1 i_1 \vet{v}^{(1)} + \sigma_2 i_2 \vet{v}^{(2)}$. With these expressions the metasurface operation can be directly visualized on the Poincar\'e sphere, as illustrated in Fig.~4. In essence, the metasurface decomposes the incident polarization state over the polarization states identified by $\vet{w}^{(1,2)}$, reverses the handedness of these states (by converting $\vet{w}^{(1,2)} \rightarrow \vet{v}^{(1,2)}$), and scales the energy content of each polarization state, according to the singular values $\sigma_{1,2}$. The location of the polarization states is connected with the parameters $\psi$ and $\theta$ that appear in the matrix decomposition reported above (Fig.~4). As expected, $\psi$ is connected to a physical rotation (a rotation in the plane of the coordinate axes identifying the $x$- and $y$-related Stokes parameters $S_{1,2}$). Instead, $\theta$ is directly connected with the ``chiral'' Stokes parameter $S_3$. In the following, we will mostly drop the use of $\theta$ and will rather define the metasurface functionality referring to the $S_3$ values of $\vet{w}^{(1,2)}$: $S_3^{(1,2)} = \pm \cos 2\theta$. 

\begin{figure}[htbp]
\begin{center}
\includegraphics{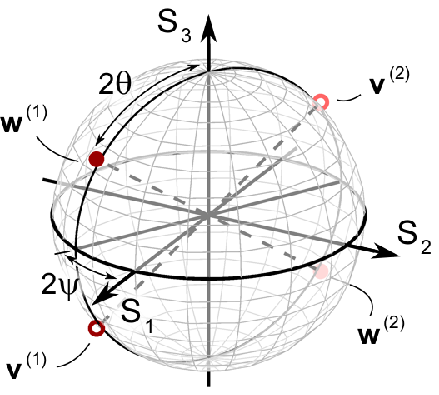}%
\end{center}
\caption{\label{fig4} Illustration of the action of a 2d-chiral metasurface on the Poincar\'e sphere. $\vet{w}^{(1,2)}$ and $\vet{v}^{(1,2)}$ identify the polarization states of, respectively, right and left singular vectors of the transmission (Jones) matrix. The metasurface operation is to map the $\vet{w}$'s into the $\vet{v}$'s, plus a rescaling described by the singular values (see text for details).}
\end{figure}
To complete the picture one should also consider the phase response of the metasurface, which is described by the parameters $\tilde{\phi}$ and $\Delta$ that appear in the decomposition. These parameters are connected to a relative and a global phase, in the following sense: $\Delta$ identifies the phase difference between the $R$ components of $\vet{w}^{(1,2)}$, since $2\Delta = \mathrm{arg}\left(w_{R}^{(1)}\right) - \mathrm{arg}\left(w_{R}^{(2)}\right)$. Instead, the phase difference between incident and transmitted field, $\phi = \mathrm{arg}\left(v_{R}^{(1)}\right) - \mathrm{arg}\left(w_{R}^{(1)}\right)$ is given by $\phi = \tilde{\phi} - 2\Delta + \pi$. Both relative and global phases are of practical interest, as they allow for independent control over the phase profile of both polarizations in polarization-dependent metasurface holograms.

The algebraic and parametric analysis performed so far is valid at each individual wavelength. More information can however be obtained by studying the spectral dispersion of the SVD parameters, whose frequency-dependent response can be eventually correlated with quantities of more direct experimental access. 
\begin{figure}[htbp]
\includegraphics{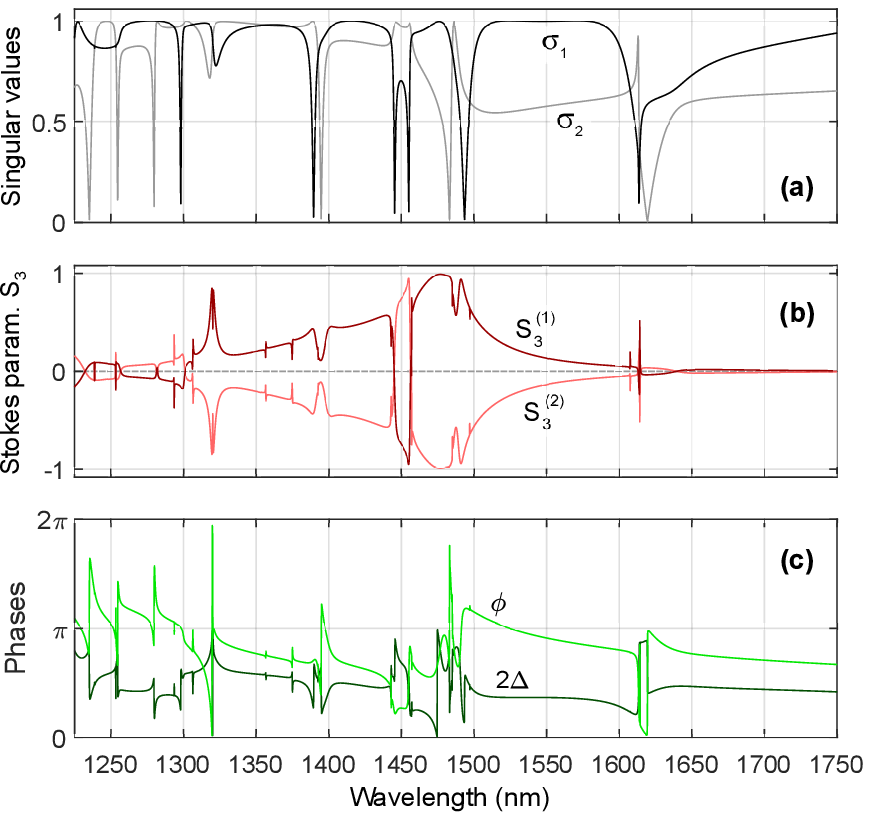}%
\caption{\label{fig5} Main quantities arising from the singular value decomposition (SVD) of the metasurface transmission matrix at normal incidence. Panel (a), singular value spectra. Note that the singular values structures recall the features observed in the transmittance, circular dichroism and electromagnetic chirality spectra (Figs.~1 and 3). Panel (b), third component of the Stokes parameter of the right singular vector. $S_3^{(1,2)}$ correspond, respectively, to singular values $\sigma_{1,2}$. Panel(c), relative phase $2 \Delta$ (i.e., phase between right singular vectors) and absolute transmission phase $\phi$. }
\end{figure}
In Fig.~5a we plot the calculated spectral dispersion of the singular values for the L-shaped hole array described in the previous sections. In this analysis, singular values are not sorted in decreasing order, rather, they are sorted such as their trend with respect to the wavelength is smooth. The spectrum reveals a structure with narrow dips, mostly occurring in pairs, which stand out of a background where both $\sigma$'s are close to 1 (for $\lambda < 1.45\ \um$) or where one $\sigma$ is close to 1 and the other is close to 0.5 ($\lambda > 1.45\ \um$). Noticeably, these dips occur at the same wavelengths where the spectra of transmission, circular dichroism, and near-field chirality also show peaks or dips. It is also interesting to notice that the dips reach zero: at those wavelengths, the metasurface does not transmit the radiation which is incident with the polarization dictated by the corresponding right singular vector. Fig.~5b reports instead the spectral dependence of the third Stokes parameter of the right singular vectors. Here it can be noticed that the $S_3^{(1,2)}$ spectra show a large peak centered around $1.47\ \um$: this feature clearly recalls the localized, broadband resonance discussed in Sect.~3. On top of that, narrow features, of widths comparable to those observed in $\sigma_{1,2}$, are present. Finer features, often occurring close to the points where $\sigma_{1} = \sigma_{2}$, have an unclear interpretation and may be connected to singular value degeneracy. In the wavelength region around $1.45\ \um$ a reversal of the signs of $S_3^{(1,2)}$ is also observed. The spectral features present on $\sigma_{1,2}$ and on $S_3^{(1,2)}$ clearly recall what noticed on the observables that have been studied in the previous sections (transmission, circular dichroism, near-field chirality): we attribute this fact to the presence of a common ground -- i.e., metasurface resonances -- standing behind all these phenomena.
We finally reported in Fig.~5c the spectra of the phases $\phi$ and $2 \Delta$, which also show a rich behavior, with rapid changes occurring close to the zeros of the singular values. In general, the spectral behavior of the SVD parameter is rather complex. While this fact, from one side, opens further challenges in view of a comprehensive interpretation, it also holds promises in view of metasurface engineering: indeed, it means that there is a wide space where the metasurface functionality can be chosen and designed.

\section{Inverse problem and metasurface minimality}

\begin{figure*}[htbp]
\centering
\includegraphics{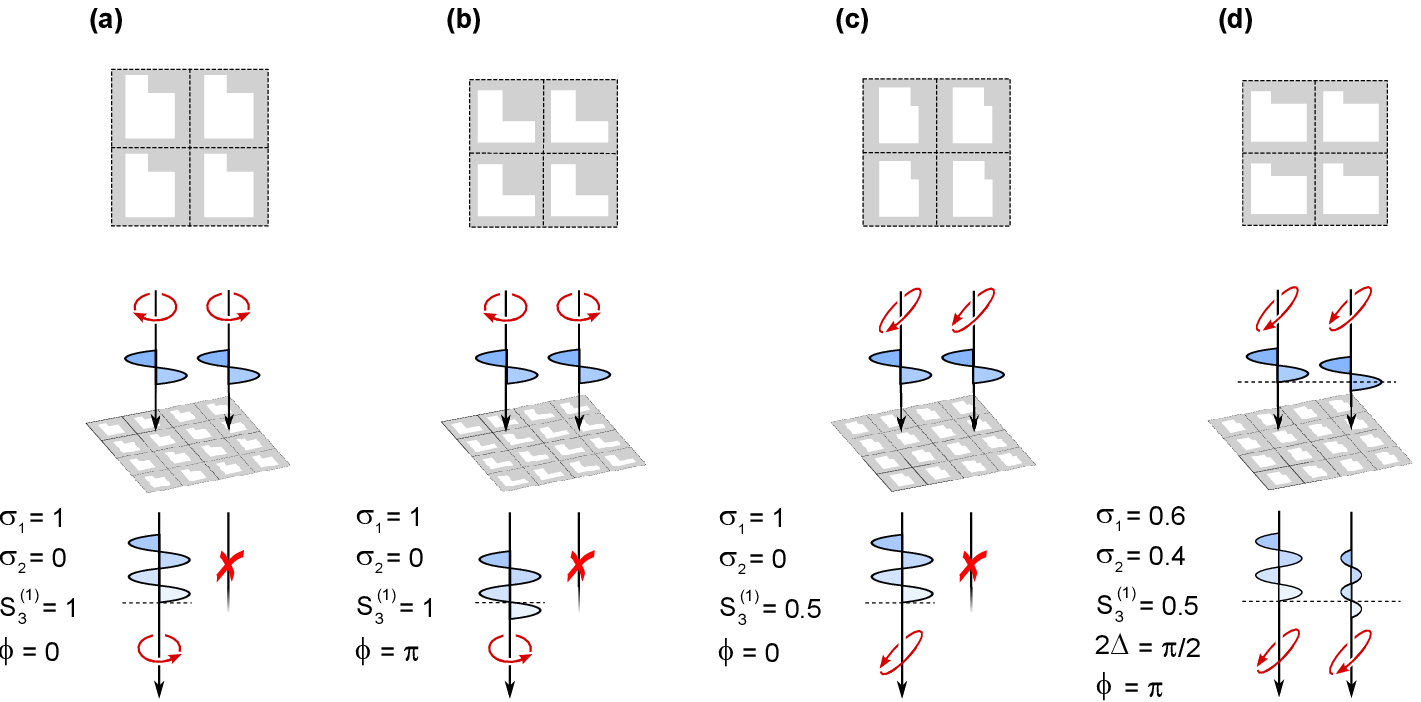}%
\caption{\label{fig6} Example of wave operations achievable with the minimal metasurface: (a) circular polarizer, (b) circular polarizer + phase delay, (c) elliptical polarizer, (d) elliptical diattenuator + phase delay. The L-shapes illustrated in the top of the figure depict the actual geometry required to implement the aforementioned operations. The exact values of the geometrical parameters ($a$, $f_1 \ldots f_4$) are given in the Supplementary Material. in correspondence to each case we also report the SVD parameters (see Sect.~\ref{sect:svd} for details). A much wider library of L-shapes that realize wave operation targets spanning the whole accessible transmission matrix space is reported in the Supplementary Material.}
\end{figure*}
Provided with the SVD formalism, and motivated by the conclusions of the previous section, an intriguing question is whether it is possible to solve the inverse problem: given a target metasurface function (i.e., a target $T$ matrix, or better a set of target parameters $\sigma_{1,2}$, $S_3^{(1)}$, $\Delta$, $\phi$, and $\psi$), does it exist a metasurface unit cell shape that induces it? We determined that appropriately shaped L-holes in a 220 nm thick dielectric slab with a refractive index $n = 3.38$ (i.e., that of GaAs at $1.55\ \um$), are capable of implementing an arbitrary $T$ matrix of the class pertinent to $M_{x,y}$ objects.  In this analysis the technologically relevant case of normal incidence is assumed. We approached the inverse problem by choosing a large set (120 combinations) of target parameters; the complete list is reported in the Supplementary Material. By running an optimization algorithm we have been able to identify, for all the targets, a specific geometry of the L-shaped hole such that the calculated $T$ matrix matches the target. It is interesting to notice that the L-shaped hole has 5 degrees of freedom ($a$, $f_1 \ldots f_4$), the same as the number of non-trivial degrees of freedom of the $T$ matrix of $M_{x,y}$ objects: $\sigma_{1,2}$, $S_3^{(1)}$, $\Delta$ and $\phi$. In this sense we have defined the metasurface as \textit{minimal}. We recall that the parameter $\psi$ is trivial in the sense defined in the previous section, since it can be targeted by simply rotating the metasurface. While the complete solution dataset is available in the Supplementary Material, we report here four exemplary cases. Three of them are of immediate and intuitive interpretation; they are illustrated in Fig.~6 a-c. The target objects are here polarizers, which are hence characterized by $\sigma_1 = 1$, $\sigma_2 = 0$. The first (a) is an ordinary circular polarizer, the second (b) is a circular polarizer that also imprints a phase delay, and the third (c) is an elliptical polarizer. More precisely, the objects  under consideration are polarizers \textit{and} handedness inverters: thanks to the special structure of the singular vectors, a $M_{x,y}$ object with $\sigma_1 = 1$ and $\sigma_2 = 0$ eliminates one of the polarization component of the incident field, and transmits the other with inverted handedness (see Fig.~4 and the related discussion). The fourth example, Fig.~6d, is an implementation of the most general type of $T$-matrices that can be targeted by means of $M_{x,y}$ objects: here, all the target parameters ($\sigma_{1,2}$, $S_3^{(1)}$, $\Delta$, $\phi$) assume non-trivial values. The input and output singular states are elliptic polarization states, there is a phase difference between the input singular states, there is a phase difference between the output singular states, and the singular values are different from each other, and different from 0 and 1. In other words the metasurface behaves as the combination of a diattenuator and a retarder acting over elliptical states. Implementing this operation by means of conventional optical elements would require a bulky stack of layers, which are here replaced by a simple, subwavelength film where the operation is fully encoded in the shape of the lithographically defined holes. One might hence envisage ultracompact and ultralight optical components acting over all the key parameters of light -- amplitude, phase and polarization. What is more, the possibility to access the full $T$-matrix parameter space by means of a simple hole-shape tuning of a fixed-thickness membrane allows to implement space-variant metasurfaces capable of performing more advanced operations with respect to what is known to date \cite{BalthasarPRL2017}, with possible applications to beam shaping, holography, and cryptography. As a final remark we highlight that the targeting procedure, which we performed at a wavelength of $1.55\ \um$ that is of direct interest to the telecommunication technology, is fully scalable, thanks to the scale invariance of Maxwell equations. 

\section{Conclusions}
In conclusion we reported about the observation of various chiro-optical phenomena occurring in a 2d-chiral patterned dielectric slab, i.e., in a patterned slab that exhibits a single mirror symmetry plane parallel to the slab itself. Transmission circular dichroism is arranged in dispersive bands, and shows the fingerprints of localized and delocalized photonic resonances. It is in particular the effect of a localized resonance that induces an intrinsic chiral response on the metasurface, with consequences on the near-field chirality that shows superchirality and other anomalous behaviors. Relying on the singular value analysis of the transmission matrix, we identified the key parameters describing in full the operation of a 2d-chiral slab. We also noticed connections between the singular value spectrum and the above cited phenomena. Finally, we showed that the L-shaped hole structure, i.e., the most intuitive 2d-chiral pattern, is also a minimal one, as it allows to implement arbitrary transmission matrices with a minimal number of parameters. This result about inverse design may open the way towards advanced space-variant metasurfaces that exploit in full the phase, amplitude and polarization degrees of freedom of light.

\section*{Funding Information}
This work was in part supported by the European Commission through the project PHENOMEN (H2020-EU-713450). 

\section*{Acknowledgments}

We acknowledge Alberto Bordin, who participated in an early stage of the project, Stefano Luin (Scuola Normale Superiore, Pisa) for support on the data fitting, Francesca Bontempi (Scuola Superiore Sant'Anna, Pisa) for ellipsometric measurements of the dielectric thin film, Sara Nocentini and Lorenzo Pattelli (LENS, Florence) for useful discussions and precious support for the spectroscopic measurements.

\section{Supplementary Material}
\subsection{Metasurface fabrication}
The metasurface samples have been fabricated starting from a GaAs wafer, where an epitaxial bilayer consisting of $\alzcq$ (1500 nm) and GaAs (220 nm) has been deposited. Several L-shaped hole arrays have been defined by means of e-beam lithography on a 8 mm $\times$ 8 mm chip (Zeiss Ultraplus SEM + Raith Multibeam lithography system; 30 keV, AR-P 6200 resist). The pattern has then been transferred on the semiconductor layer by means of ICP-RIE with $\mathrm{Cl}_2$/$B\mathrm{Cl}_3$/$\mathrm{Ar}$ gas mixture (Sentech). To obtain the patterned membrane, the chip has been etched from the back side with a sequence of wet etching steps. First, a fast etching ($\mathrm{H}_2\mathrm{SO}_4$\,:\,$\mathrm{H}_2\mathrm{O}_2$\,:\,$\mathrm{H}_2\mathrm{O}$, 1:8:1) with $\approx 10\ \micron$ per minute etch rate is employed to thin the substrate down to $\approx 50\ \micron$. 
\begin{figure*}
\centering
\includegraphics{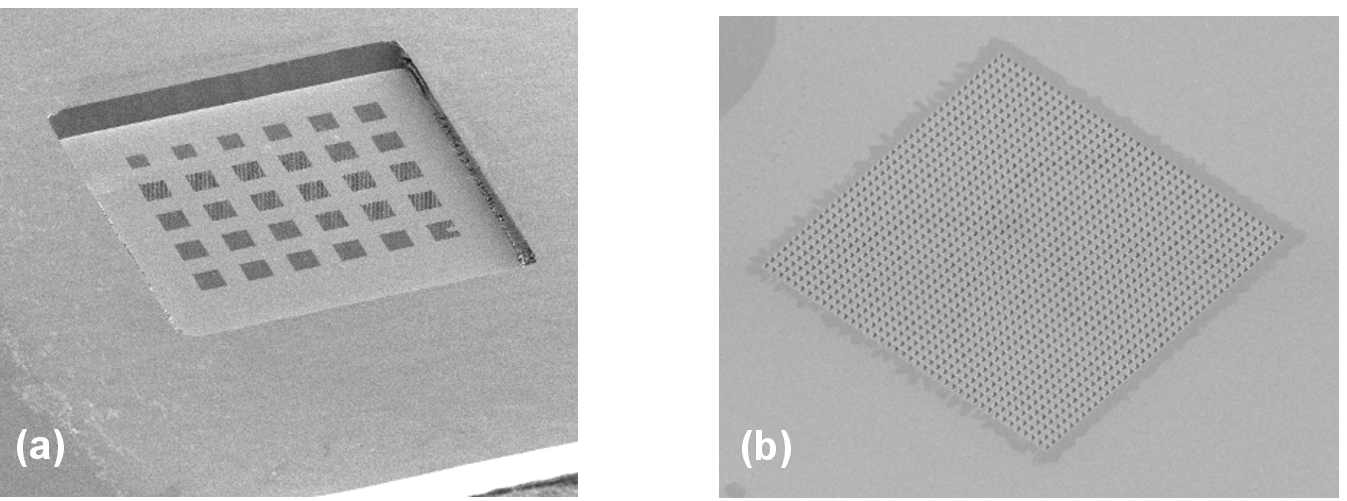}
\caption{Scanning electron microscope images of the fabricated metasurface sample. (a) An $~ 800\ \micron \times 800\ \micron$ sized membrane containing a group of L-shaped hole arrays. The chip is here observed from the bottom side. (b) A single L-shaped hole array, of size $~ 50\ \micron \times 50\ \micron$.}
\end{figure*}
A resist mask (S1818) is employed in order to protect $\approx 1\ \mathrm{mm}$ of chip on the edges, thus allowing for chip handling. Second, a slow etching (3:1 solution of citric acid in 30\% hydrogen peroxide; citric acid is prepared in 1:1 weight ratio with water) having $\approx 1\ \micron$ per minute etch rate is employed to remove all the remaining GaAs substrate. This solution is selective and stops when the $\alzcq$ layer is reached. In this phase a second resist mask is employed, in order to realize membranes of $\approx 800\ \micron \times 800\ \micron$ size. Finally, an HF bath (50\% concentration) is employed to remove the $\alzcq$ layer and obtain the frame-supported patterned membrane (Fig.~7). During the first two wet etching steps the chip is mounted on a glass slide by means of S1818 resist.

\subsection{Numerical calculations}
The transmission matrix (Jones matrix) of the metasurface has been calcuated by means of a  \textsc{MATLAB} software that implements the rigorous coupled wave analysis scattering matrix method (RCWA) following the formalism of \cite{Whittaker1999, Liscidini2008, Li1996, Lalanne1996}. The theoretical spectra in Figs.~1 and 2 have been smoothed by convolution with a 2 nm wide Gaussian in order to better  compare them with the experimental ones.

The field plots reported in Fig.~2 of the main text have instead been made with the finite-element software \textsc{COMSOL}.

\subsection{Empty lattice band dispersion}

To understand the nature of the dispersive features observed in the angularly-resolved circular dichroism spectra reported in the main text, we performed an empty-lattice band dispersion calculation. First we identified the guided modes supported by an unpatterned dielectric slab, whose thickness and refractive index is the same as that of the gallium arsenide slab employed to fabricate the metasurface. The modal effective refractive indices ($n_{\mathit{eff}}$) are reported in Fig.~8a: in the wavelength range of interest the slab supports the fundamental TE and TM modes ($\mathrm{TE}_0$ and $\mathrm{TM}_0$) and, below the cutoff wavelength of $\approx 1.4\ \um$, also the first-order modes. 
\begin{figure*}
\centering
\includegraphics{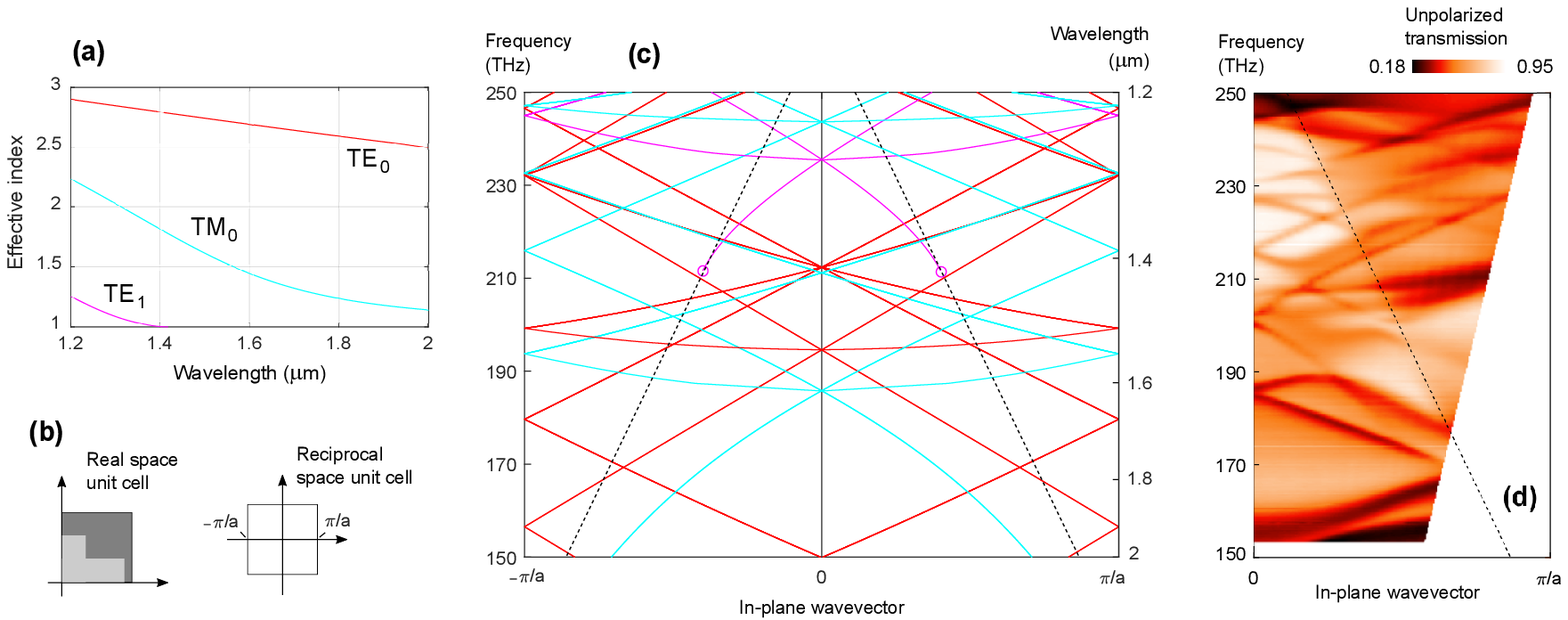}
\caption{Empty lattice band dispersion from the guided modes of a dielectric slab. Panel (a), effective index dispersion for the unpatterned slab guided modes. Panel (b), real space unit cell (i.e., a square unit cell enclosing the L-shaped hole) and reciprocal space unit cell (i.e., first Brillouin zone). Panel (c), band dispersion originating from the mode folding. The circles at $\approx 1.4 \um$ indicate the $\mathrm{TE}_1$ cutoff; the diagonal dashed line represent the diffractin threshold (i.e., the folded air light cone). Panel (d), experimental unpolarized angle-resolved transmission.}
\end{figure*}
As the $\mathrm{TM}_1$ mode has a very low value of $n_{\mathit{eff}}$, the plot reports only the $\mathrm{TE}_1$ mode. Second, we folded the guided modes in the first Brillouin zone, according to the zero-order perturbation theory summarized by the equation
\[
\left| \vet k_B + \vet g \right| = \beta
\]
where $\vet k_B$ is the Bloch wavevector, $\vet g$ is a reciprocal lattice vector, and $\beta$ is the guided mode propagation constant (i.e., $\beta = 2 \pi n_{\mathit{eff}} / \lambda$) \cite{CollinRPP2014}. Consistently with the experiments, the Bloch wavevector is varied within the first Brillouin zone along the high-symmetry direction indicated in Fig.~8b. The resulting mode dispersions are reported in Fig.~8c; notice that they appear symmetric since no symmetry-breaking mechanisms are encoded in this simple model. Despite the simplicity of the model, a comparison between theory and experiment (i.e., between Figs.~8c-d) shows that the overall trend of the features observed in the experiment are appropriately grasped by the model. This confirms that the dispersive bands observed in the transmission spectra, and hence in the circular dichroism spectra, originate from a guided-mode resonance mechanism.

\subsection{Demonstration of the algebraic structure of \textit{T}-matrix singular value decomposition}

In Sect.~5 of the main text we reported that the $T$-matrix of a metasurface belonging to the $M_{x,y}$ symmetry class can be expressed by means of singular value decomposition (SVD) in a specific form. We briefly outline here the demonstration of that result. First, we recall that any complex matrix $M$ can be decomposed in the form $M = U_1 \Sigma U_2^{\dagger}$, where $U_{1,2}$ are unitary matrices whose columns are the (right and left) singular vectors of $M$, and $\Sigma$ is a diagonal matrix whose entries are real non-negative numbers (the singular values) \cite{HornJohnson}. The SVD plays an important role in the theory of linear multiport lossy optical components, since the singular values are the minimum and maximum absorption levels that the component can implement \cite{Ge2017}. 

First observation is that the SVD of $T_L$, which is a symmetric matrix by hypothesis ($T_L = T_L^t$), can be cast to a special form. Indeed, from the above definition it turns out that one can write $T_L = \bar{U} \Sigma U^{\dagger}$, where the bar indicate the element-wise matrix complex conjugate. Matrix $T_C$ is obtained from $T_L$ by the linear-to-circular basis conversion matrix\footnote{In the notation we employ, circular polarization basis is sorted according to (RCP,LCP). RCP is defined as following: at fixed position in space, the electric vector rotates in the sense dictated by the fingers of a right hand whose thumb is pointing along the propagation direction. Moreover, we use the convention $e^{-i \omega t}$. }, i.e., $T_C = \Lambda^{\dagger} T_L \Lambda$, where $\Lambda = \frac{1}{\sqrt{2}} \left( \begin{array}{cc} 1 & 1 \\ i & -i  \end{array} \right)$. Looking for a SVD of $T_C$, i.e., for an expression $T_C = V \Sigma W^{\dagger}$, it turns out immediately that $V$ and $W$ must satisfy $\bar{V}W = \left( \begin{array}{cc} 0 & 1 \\ 1 & 0 \end{array} \right)$. By inspection we obtained the decomposition of $V$ and $W$ in more elementary matrices that is reported in the main text.

\subsection{Complete dataset of metasurface shape parameters needed to target arbitrary functions}

In Sect.~6 of the main text we claimed that, by appropriately shaping the L-shaped hole, an arbitrary $T$-matrix can be obtained. The problem has been approached numerically by defining an error function that quantifies the distance of the $T$-matrix of a metasurface with certain geometrical parameters (i.e., $T (a, f_1 \ldots f_4) $) from the target $T$-matrix (i.e., $T_{\mathit{targ}}$). According to the matrix parametrization given in Sect.~5, the error function is defined as
\begin{equation*}
\begin{split}
\mathit{error} = & \left( \sigma_1 - \sigma_{1,\mathit{targ}} \right)^2 + \left( \sigma_2 - \sigma_{2,\mathit{targ}} \right)^2 + \\
& + \left( \left( S_3^{(1)} - S_{3,\mathit{targ}}^{(1)} \right)/2 \right)^2  +\\
& + \sin^2 \left( \frac{2 \Delta - 2\Delta_{\mathit{targ}} }{2} \right) + 
\sin^2 \left( \frac{\phi - \phi_{\mathit{targ}} }{2} \right)
\end{split}
\end{equation*}
and it is used in an optimization process that employs the \texttt{fmincon} function of the Optimization Toolbox of \textsc{MATLAB}. Notice that we employed $2\Delta$ rather than $\Delta$, since it is the former which has physical significance (see the main text). The constraints on the geometric parameters are the following: $a \in [0.8,1.5]\ \um$, $f_{1 \ldots 4} \in [0.2,0.9]$, $f_3 < f_1$, $f_4 < f_2$. For each target, several optimization processes with random starting points were needed in order to get a solution within an acceptable error level ($\mathit{error} < 0.02$).

While approaching the process of targeting arbitrary metasurface functions, one might extract random combinations of target $\sigma_{1,\mathit{targ}}$, $\sigma_{2,\mathit{targ}}$, $S_{3,\mathit{targ}}^{(1)}$, $\Delta_{\mathit{targ}}$ and $\phi_{\mathit{targ}}$ and solve the inverse problem. However, further analysis on the meaning of these parameters suggests that certain subsets deserve more relevance than others. For instance, the combinations where $\sigma_{1,\mathit{targ}} = 1$ and $\sigma_{2,\mathit{targ}} = 1$ can be excluded from the present analysis. Indeed, this request means that the $T$-matrix must be unitary, and that its eigenvectors must be orthogonal. However, following \cite{MenzelPRA2010}, the $T$-matrix of an $M_{x,y}$ object has corotating eigenvectors; merging this requirement to that of eigenvector orthogonality means that the eigenvectors must be linearly polarized. This observation implies that there is no need to employ an L-shaped structure; a simpler rotated rectangular or elliptical hole in a dielectric slab (or dielectric post on a substrate) will be sufficient to implement the required operation (see, for instance, the Supplementary Information of \cite{ArbabiNatNano2015}). For this reason we will not consider the case $\sigma_{1,\mathit{targ}} = \sigma_{2,\mathit{targ}} = 1$.
Another point to be observed is that it is sufficient to consider positive values of $S_{3,\mathit{targ}}^{(1)}$. Indeed, given a structure with geometrical parameters $[ a, f_1, f_2, f_3, f_4 ]$ whose SVD parameters are $[ \sigma_{1}, \sigma_{2}, \Delta , \phi, S_3^{(1)}]$, the structure obtained by exchanging $f_1 \leftrightarrow f_3$, $f_2 \leftrightarrow f_4$ (i.e., the enantiomer), will have the sign of $S_{3}^{(1)}$ inverted. 

A first subset that deserve particular interest is that where one of the singular values is zero: $\sigma_{2,\mathit{targ}} = 0$; this set contains polarizers. It should be noticed that in this subset the parameters $\Delta$ is irrelevant, since in the final expression of the transmission matrix ($T_C = V \Sigma W\dagger $) it only appears in the term $e^{-i(2\Delta - \tilde{\phi})} = e^{-i(\phi+\pi)}$. We chose 24 combinations of target parameters within this set, that are reported in the table of Fig.~9. In correspondence to each target, the table reports the values of the geometrical parameters that solve the inverse problem. In the table of Fig.~9 we labeled with the letters a-c the first three cases considered in Fig.~6 of the main text. 

\begin{figure*}
\begin{center}
\includegraphics[width = 11 cm]{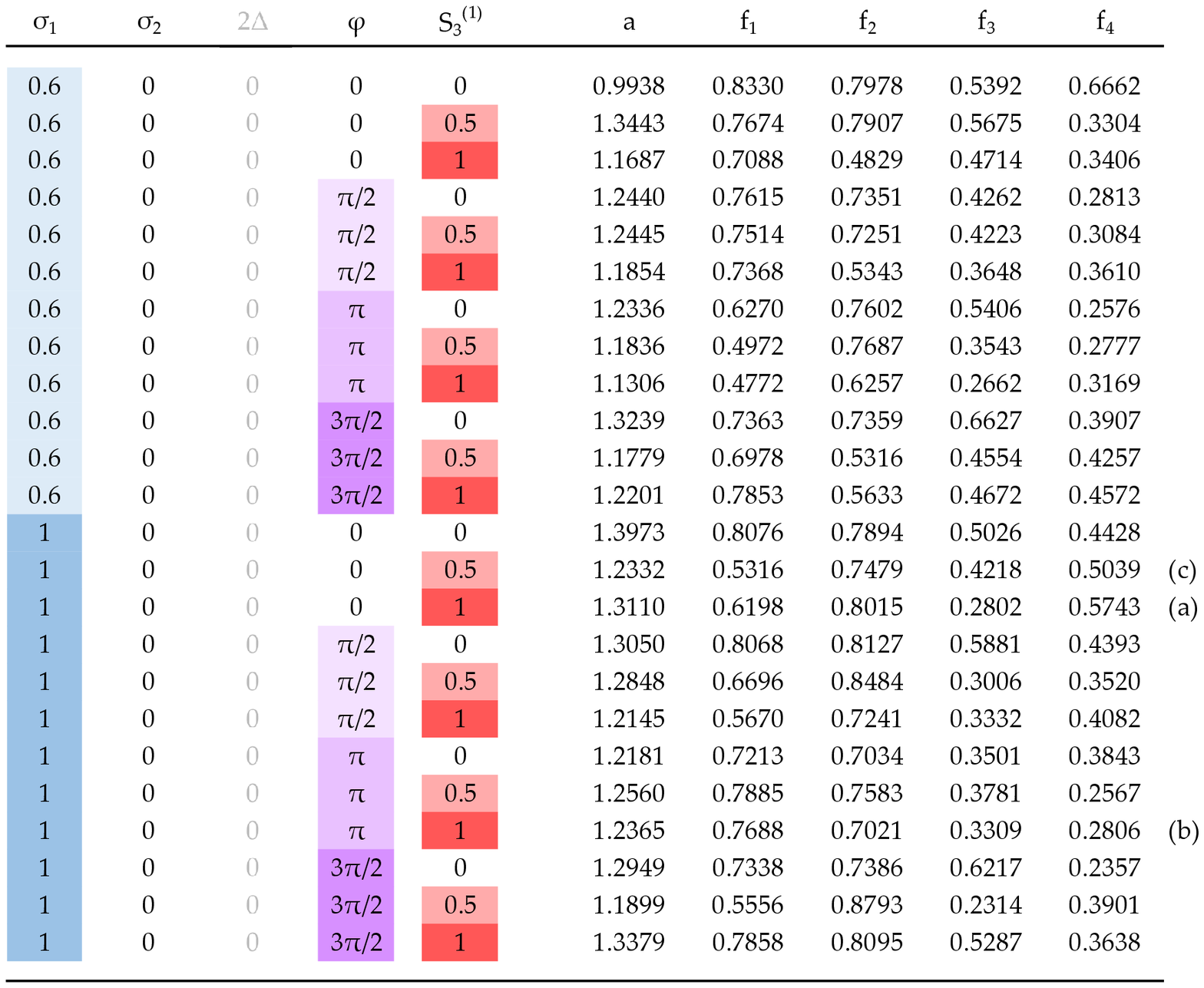}
\end{center}
\caption{Target parameters and corresponding geometrical parameters that solve the inverse problem. In the present subset, where $\sigma_2 = 0$, the value of $2 \Delta$ is irrelevant (see text). The lines marked with a letter are the cases reported in Fig.~6 of the main text.}
\end{figure*}

The second subset that has been considered is that where both singular values are non-zero, but not simultaneously equal to unity. Within this subset we chose 96 target combinations, by allowing the target parameters to assume the values $\sigma_1 = 0.6, 1$; $\sigma_2 = 0.4$; $2 \Delta = 0, \pi/4, \pi/2, 3 \pi/4 $; $\phi = 0, \pi /2, \pi, 3 \pi /2$; $S_3^{(1)} = 0, 0.5, 1$. We have limited the range of $2 \Delta$ within the interval $[0, \pi]$ since it can be shown that a certain pair of $2 \Delta$ and $\phi$ leads to the same $T$-matrix obtained  by employing $2 \Delta + \pi$ and $\phi + \pi$. The inverse problem has been solved successfully for all the 96 combinations. However, an illustrative subset is reported in the table of Fig.~10, where a random sampling of the complete dataset has been listed. Here, the letter ``d'' labels the fourth case considered in Fig.~6 of the main text. In the data of the tables of Fig.~9 and 10 there is not immediate evidence of a correlation between the target parameters and the geometrical parameters; however, more refined data analysis techniques and finer-grained parameter sweeping may reveal more details of the physics behind the inverse problem. However, the study of these aspects goes beyond the aims of the present work and may be the topic of future research.

\begin{figure*}
\begin{center}
\includegraphics[width = 11 cm]{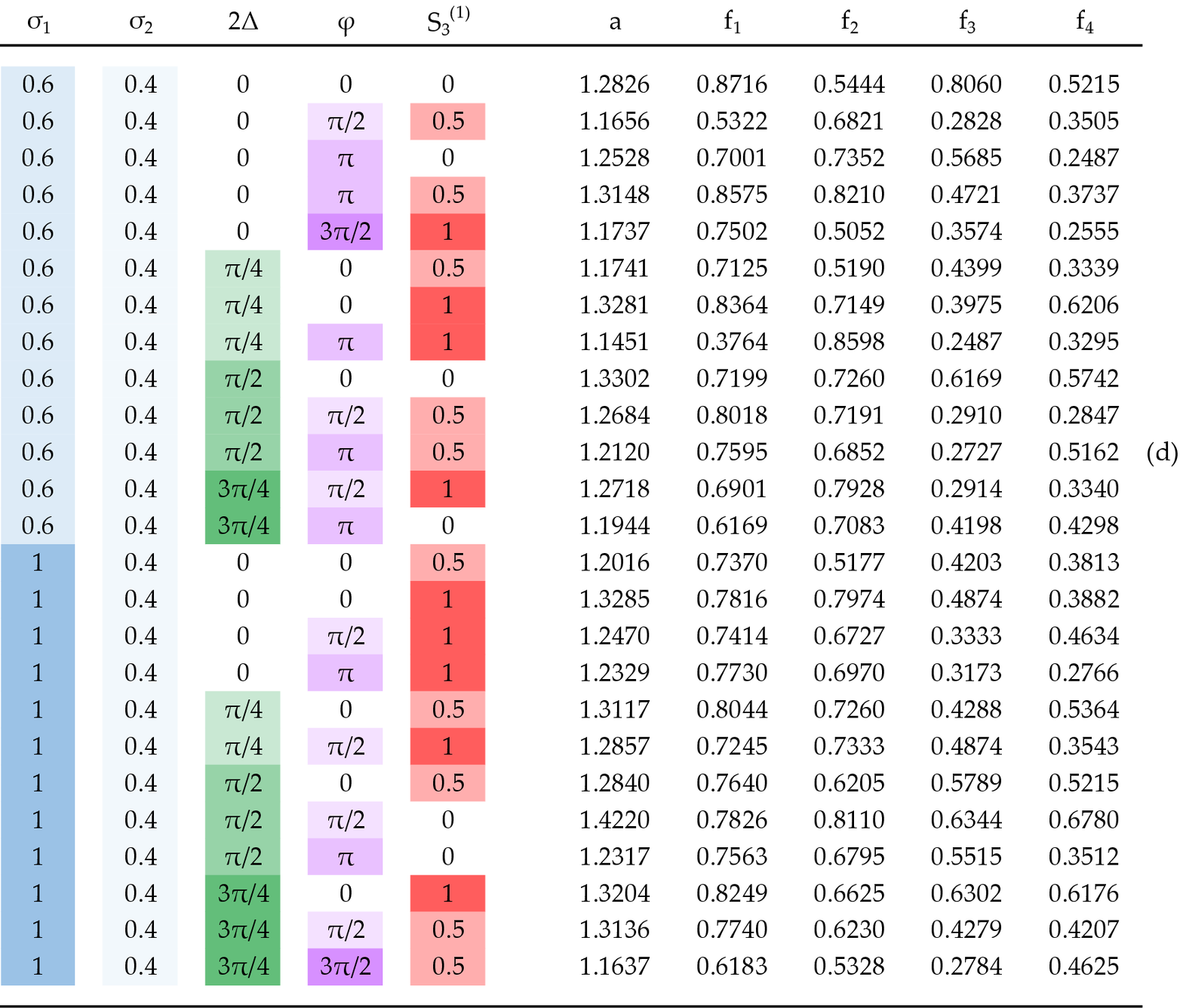}
\end{center}
\caption{Target parameters and corresponding geometrical parameters that solve the inverse problem. This table is an exemplary selection of a wider table, available upon request, that contains 96 combinations of target parameters and the corresponding geometrical parameters. The line marked with a letter is case ``d'' of the Fig.~6 of the main text.}
\end{figure*}

\bibliography{ChiralBands}

\end{document}